\documentclass[amsmath,aps,showpacs,a4paper,10pt]{revtex4}
 \usepackage{epsf}
 \usepackage{graphicx}    % Include figure files

\begin{document}

 \newcommand{\be}[1]{\begin{equation}\label{#1}}
 \newcommand{\ee}{\end{equation}}
 \newcommand{\bea}{\begin{eqnarray}}
 \newcommand{\eea}{\end{eqnarray}}
 \def\disp{\displaystyle}

 \def\gsim{ \lower .75ex \hbox{$\sim$} \llap{\raise .27ex \hbox{$>$}} }
 \def\lsim{ \lower .75ex \hbox{$\sim$} \llap{\raise .27ex \hbox{$<$}} }

 \begin{titlepage}

 \begin{flushright}
 arXiv:1109.6107
 \end{flushright}

 \title{\Large \bf Dynamics of Teleparallel Dark Energy}

 \author{Hao~Wei\,}
 \email[\,email address:\ ]{haowei@bit.edu.cn}
 \affiliation{School of Physics, Beijing Institute
 of Technology, Beijing 100081, China}

 \begin{abstract}\vspace{1cm}
 \centerline{\bf ABSTRACT}
 Recently, Geng~{\it et~al.} proposed to allow a non-minimal
 coupling between quintessence and gravity in the framework
 of teleparallel gravity, motivated by the similar one in the
 framework of General Relativity (GR). They found that this
 non-minimally coupled quintessence in the framework of
 teleparallel gravity has a richer structure, and named it
 ``teleparallel dark energy''. In the present work, we note
 that there might be a deep and unknown connection between
 teleparallel dark energy and Elko spinor dark energy.
 Motivated by this observation and the previous results of
 Elko spinor dark energy, we try to study the dynamics of
 teleparallel dark energy. We find that there exist only some
 dark-energy-dominated de~Sitter attractors. Unfortunately,
 no scaling attractor has been found, even when we allow the
 possible interaction between teleparallel dark energy and
 matter. However, we note that $w$ at the critical points is
 in agreement with observations (in particular, the fact that
 $w=-1$ independently of $\xi$ is a great advantage).
 \end{abstract}


 \maketitle

 \end{titlepage}

 \renewcommand{\baselinestretch}{1.1}

%============================= section 1 ===================================

\section{Introduction}\label{sec1}

Since the striking discovery of the current accelerated
 expansion in 1998, it has been one of the most active fields
 in modern cosmology~\cite{r1,r2,r3}. Besides the cosmological
 constant, the simplest candidate of dark energy is the
 well-known quintessence, which is described by a canonical
 scalar field $\phi$ in the framework of General Relativity
 (GR). The relevant action reads~\cite{r1}
 \be{eq1}
 {\cal S}=\int d^4 x\sqrt{-g}\left[\frac{R}{2\kappa^2}+
 \frac{1}{2}\partial_\mu \phi \partial^\mu \phi-V(\phi)\right]
 +{\cal S}_m\,,
 \ee
 where $R$ is the Ricci scalar; ${\cal S}_m$ is the matter
 action; $\kappa^2\equiv 8\pi G$; we use the metric signature
 convention $(+,-,-,-)$ throughout. Considering a spatially
 flat Friedmann-Robertson-Walker (FRW) universe
 and a homogeneous scalar field $\phi$, the corresponding
 Friedmann equation and Raychaudhuri equation are given by
 \bea
 &&H^2=\frac{\kappa^2}{3}\rho_{tot}=\frac{\kappa^2}{3}
 \left(\rho_\phi+\rho_m\right)\,,\label{eq2}\\
 &&\dot{H}=-\frac{\kappa^2}{2}\left(p_{tot}+\rho_{tot}\right)
 =-\frac{\kappa^2}{2}\left(p_\phi+\rho_\phi+p_m+
 \rho_m\right)\,,\label{eq3}
 \eea
 where $H\equiv\dot{a}/a$ is the Hubble parameter; $a$ is the
 scale factor; a dot denotes the derivatives with respect to
 cosmic time $t$; $p_m$ and $\rho_m$ are the pressure and
 energy density of background matter, respectively. In this
 work, we assume that the background matter is described by
 a perfect fluid with barotropic equation-of-state parameter
 (EoS), namely
 \be{eq4}
 p_m=w_m\rho_m\equiv (\gamma-1)\rho_m\,,
 \ee
 where the so-called barotropic index $\gamma$ is a positive
 constant. In particular, $\gamma=1$ and $4/3$ correspond to
 dust matter and radiation, respectively. As is well known,
 the pressure and energy density of quintessence are given by
 \be{eq5}
 p_\phi=\frac{1}{2}\dot{\phi}^2-V(\phi)\,,~~~~~~~
 \rho_\phi=\frac{1}{2}\dot{\phi}^2+V(\phi)\,,
 \ee
 where $V(\phi)$ is the potential. The quintessence has been
 extensively discussed in the literature, and we refer to
 e.g.~\cite{r1} for some comprehensive reviews.

As is well known, in the literature one can generalize
 quintessence by including a non-minimal coupling between
 quintessence and gravity. The relevant action
 reads~\cite{r1,r4,r5}
 \be{eq6}
 {\cal S}=\int d^4 x\sqrt{-g}\left[\frac{R}{2\kappa^2}+
 \frac{1}{2}\left(\partial_\mu \phi \partial^\mu \phi+
 \xi R\phi^2\right)-V(\phi)\right]+{\cal S}_m\,,
 \ee
 where $\xi$ is a constant measuring the non-minimal coupling.
 In this case, the corresponding Friedmann equation and
 Raychaudhuri equation are the same as Eqs.~(\ref{eq2})
 and~(\ref{eq3}), while the effective pressure and energy
 density of the non-minimally coupled quintessence (sometimes
 called ``extended quintessence'' in the literature) are
 changed to~\cite{r1,r4,r5}
 \bea
 &&p_\phi=\frac{1}{2}(1+4\xi)\dot{\phi}^2-V+
 2\xi(1+6\xi)\dot{H}\phi^2-2\xi H\phi\dot{\phi}
 -2\xi\phi V_{,\phi}+3\xi(1+8\xi)H^2 \phi^2\,,\label{eq7}\\
 &&\rho_\phi=\frac{1}{2}\dot{\phi}^2+V-6\xi H\phi\dot{\phi}
 -3\xi H^2 \phi^2\,,\label{eq8}
 \eea
 where $V_{,\phi}\equiv dV/d\phi\,$, and we have used the
 equation of motion
 \be{eq9}
 \ddot{\phi}+3H\dot{\phi}-\xi R\phi+V_{,\phi}=0\,,
 \ee
 which is equivalent to the energy conservation equation
 $\dot{\rho}_\phi+3H(\rho_\phi+p_\phi)=0$ in fact (it is worth
 noting that $R=6(\dot{H}+2H^2)$ in a spatially flat FRW
 universe). We refer to e.g.~\cite{r1,r2,r3,r4,r5,r32} for
 details.

Recently, the so-called teleparallel gravity originally
 proposed by Einstein~\cite{r6,r7} and its generalization,
 namely the so-called $f(T)$ theory~\cite{r8,r9}, attracted
 much attention in the community. In teleparallel gravity,
 the Weitzenb\"ock connection is used, rather than the
 Levi-Civita connection which is used in GR.
 Following~\cite{r6,r7,r8,r9}, here we briefly review the key
 ingredients of teleparallel gravity. The orthonormal tetrad
 components $e_i(x^\mu)$ relate to the metric through
 \be{eq10}
 g_{\mu\nu}=\eta_{ij}e_\mu^i e_\nu^j\,,
 \ee
 where Latin $i$, $j$ are indices running over 0, 1, 2, 3 for
 the tangent space of the manifold, and Greek $\mu$,~$\nu$ are
 the coordinate indices on the manifold, also running over 0,
 1, 2, 3. In teleparallel gravity, the relevant action is
 \be{eq11}
 {\cal S}=\frac{1}{2\kappa^2}\int d^4 x\,|e|\,T+{\cal S}_m\,,
 \ee
 where $|e|={\rm det}\,(e_\mu^i)=\sqrt{-g}\,$. The torsion
 scalar $T$ is given by
 \be{eq12}
 T\equiv{S_\rho}^{\mu\nu}\,{T^\rho}_{\mu\nu}\,,
 \ee
 where
 \bea
 {T^\rho}_{\mu\nu} &\equiv &e^\rho_i\left(\partial_\mu e^i_\nu
 -\partial_\nu e^i_\mu\right)\,,\label{eq13}\\
 {K^{\mu\nu}}_\rho &\equiv &-\frac{1}{2}\left({T^{\mu\nu}}_\rho
 -{T^{\nu\mu}}_\rho-{T_\rho}^{\mu\nu}\right)\,,\label{eq14}\\
 {S_\rho}^{\mu\nu} &\equiv &\frac{1}{2}\left({K^{\mu\nu}}_\rho
 +\delta^\mu_\rho {T^{\theta\nu}}_\theta-
 \delta^\nu_\rho {T^{\theta\mu}}_\theta\right)\,.\label{eq15}
 \eea
 For a spatially flat FRW universe, it is easy to find that
 \be{eq16}
 T=-6H^2.
 \ee
 So, one can use $T$ and $H$ interchangeably. As is well
 known, the FRW universe described by action~(\ref{eq11}) is
 completely equivalent to a matter-dominated universe in the
 framework of GR, and hence cannot be accelerated. In the
 literature, there are two ways out. In analogy to $f(R)$
 theory, the first approach is to generalize teleparallel
 gravity to $f(T)$ theory by modifying action~(\ref{eq11}) to
 \be{eq17}
 {\cal S}=
 \frac{1}{2\kappa^2}\int d^4 x\,|e|\,f(T)+{\cal S}_m\,,
 \ee
 where $f(T)$ is a function of the torsion scalar $T$.
 Recently, $f(T)$ theory attracted much attention in the community,
 and we refer to e.g.~\cite{r8,r9,r10,r11,r12,r13,r14,r33} for some
 relevant works. Obviously, the second approach is to directly
 add dark energy into teleparallel gravity. Of course, the simplest
 candidate of dark energy is still quintessence, and the
 relevant action is given by
 \be{eq18}
 {\cal S}=\int d^4 x\,|e|\left[\frac{T}{2\kappa^2}
 +\frac{1}{2}\partial_\mu \phi \partial^\mu \phi-V(\phi)\right]
 +{\cal S}_m\,.
 \ee
 However, one can easily find that dark energy in the framework
 of teleparallel gravity is completely identical to the one in
 the framework of GR, and hence there is nothing new. Very
 recently, motivated by the similar one in the framework of GR,
 Geng~{\it et~al.}~\cite{r15} proposed to modify action~(\ref{eq18})
 by including a non-minimal coupling between quintessence and
 gravity in the framework of teleparallel gravity, namely
 \be{eq19}
 {\cal S}=\int d^4 x\,|e|\left[\frac{T}{2\kappa^2}
 +\frac{1}{2}\left(\partial_\mu \phi \partial^\mu \phi+
 \xi T\phi^2\right)-V(\phi)\right]+{\cal S}_m\,.
 \ee
 They found that this non-minimally coupled quintessence in the
 framework of teleparallel gravity has a richer structure, and
 named it ``teleparallel dark energy''~\cite{r15}. The corresponding
 Friedmann equation and Raychaudhuri equation are the same as
 Eqs.~(\ref{eq2}) and~(\ref{eq3}), while the effective pressure
 and energy density of teleparallel dark energy are given
 by~\cite{r15}
 \bea
 &&p_\phi=\frac{1}{2}\dot{\phi}^2-V(\phi)+4\xi H\phi\dot{\phi}
 +\xi\left(3H^2+2\dot{H}\right)\phi^2\,,\label{eq20}\\
 &&\rho_\phi=\frac{1}{2}\dot{\phi}^2+V(\phi)
 -3\xi H^2\phi^2\,.\label{eq21}
 \eea
 The equation of motion reads
 \be{eq22}
 \ddot{\phi}+3H\dot{\phi}-\xi T\phi+V_{,\phi}=0\,,
 \ee
 which is equivalent to the energy conservation equation
 $\dot{\rho}_\phi+3H(\rho_\phi+p_\phi)=0$ in fact. Obviously,
 from Eqs.~(\ref{eq20}) and~(\ref{eq21}), the EoS
 of teleparallel dark energy $w_\phi=p_\phi/\rho_\phi$ can
 cross the phantom divide $w=-1$.

In the present work, we are interested in teleparallel dark
 energy because we note that it is reminiscent of the so-called
 Elko spinor dark energy~\cite{r16,r17,r18}. The Elko spinor
 was originally proposed by Ahluwalia and Grumiller~\cite{r19},
 which is a spin one half field with mass dimension one. Unlike
 the standard fields which obey $(CPT)^2=1$, the Elko spinor is
 a non-standard spinor according to the Wigner
 classification~\cite{r20} and obeys the unusual property
 $(CPT)^2=-1$ instead. In fact, the Elko spinor fields
 (together with Majorana spinor fields) belong to a wider class
 of spinor fields, i.e., the so-called flagpole spinor fields,
 according to the Lounesto general classification of all spinor
 fields~\cite{r21,r34}. The effective pressure and energy
 density of Elko spinor dark energy are given by~\cite{r17,r18}
 \bea
 && p_\phi=\frac{1}{2}\dot{\phi}^2-V(\phi)
 -\frac{3}{8}H^2\phi^2-\frac{1}{4}\dot{H}\phi^2-
 \frac{1}{2}H\phi\dot{\phi}\,,\label{eq23}\\
 && \rho_\phi=\frac{1}{2}\dot{\phi}^2+V(\phi)+
 \frac{3}{8}H^2\phi^2\,.\label{eq24}
 \eea
 We strikingly find that when $\xi=-1/8$, Eqs.~(\ref{eq20})
 and~(\ref{eq21}) become identical to Eqs.~(\ref{eq23})
 and~(\ref{eq24}). If it is not an accident, this notable
 observation might hint a deep and unknown connection between
 teleparallel dark energy and Elko spinor dark energy. In
 particular, one might consider the deep relation between
 spinor and torsion. However, this is out of the scope of the
 present work, and we leave it as an open question. Here, we
 instead focus on another issue concerning the cosmological
 coincidence problem. As is shown in~\cite{r18,r22} by using
 the dynamical system method, Elko spinor dark energy is
 plagued with the cosmological coincidence problem. Noting the
 aforementioned connection between teleparallel dark energy
 and Elko spinor dark energy, it is very natural to ask
 whether or not the cosmological coincidence problem could be
 alleviated in teleparallel dark energy which has an extra
 free model parameter $\xi$. This is our main goal of the
 present work.

Noting that if $\xi=0$, teleparallel dark energy reduces to
 the ordinary quintessence (which is a very trivial case),
 we assume $\xi\not=0$ throughout this work.

%============================= section 2 ===================================

\section{Dynamical system of teleparallel dark energy}\label{sec2}

As is well known, the observational data tell us that we are
 living in an epoch in which the dark energy density and the
 matter energy density are comparable~\cite{r1}. However, since
 the densities of dark energy and matter scale differently with
 the expansion of our universe, there should be some kinds of
 fine-tunings. This is the well-known cosmological coincidence
 problem~\cite{r1}. Usually, this problem can be alleviated in
 most dark energy models via the method of scaling solution.
 In fact, since the nature of both dark energy and dark matter
 is still unknown, there is no physical argument to exclude
 the possible interaction between them. On the contrary, some
 observational evidences of the interaction in dark sector have
 been found recently (see e.g.~\cite{r23,r24}). If there is a
 possible interaction between dark energy and matter, their
 evolution equations could be rewritten as a dynamical
 system~\cite{r25} (see also e.g.~\cite{r26,r27,r28,r29,r30}).
 There might be some scaling attractors in this dynamical
 system, and both the fractional densities of dark energy and
 matter are non-vanishing constants over there. The universe
 will eventually enter these scaling attractors regardless of
 the initial conditions, and hence the cosmological coincidence
 problem could be alleviated without fine-tunings. This method
 works fairly well in most dark energy models (especially the
 scalar field models). However, in a few of dark energy models
 this method fails because there is no scaling attractor being
 found. As mentioned above, Elko spinor dark energy model is
 an example of failures~\cite{r18,r22}. In the present work,
 we hope that teleparallel dark energy could avoid this fate
 with the help of the extra free model parameter $\xi$,
 although there is a deep connection between teleparallel dark
 energy and Elko spinor dark energy as mentioned above.

To be general, we assume that teleparallel dark energy and
 background matter interact through a coupling term $Q$,
 according to
 \bea
 &&\dot{\rho}_\phi+3H\left(\rho_\phi
 +p_\phi\right)=-Q\,,\label{eq25}\\
 &&\dot{\rho}_m+3H\left(\rho_m+p_m\right)=Q\,,\label{eq26}
 \eea
 which preserves the total energy conservation equation
 $\dot{\rho}_{tot}+3H\left(\rho_{tot}+p_{tot}\right)=0$.
 Obviously, $Q=0$ means that there is {\em no} interaction
 between teleparallel dark energy and background matter. Also
 to be general, we assume that the background matter could be
 characterized by Eq.~(\ref{eq4}). Following
 e.g.~\cite{r25,r26,r27,r28,r29,r30}, we introduce the
 following dimensionless variables
 \be{eq27}
 x\equiv\frac{\kappa\dot{\phi}}{\sqrt{6}H}\,,~~~~~~~
 y\equiv\frac{\kappa\sqrt{V}}{\sqrt{3}H}\,,~~~~~~~
 u\equiv\kappa\phi\,,~~~~~~~
 v\equiv\frac{\kappa\sqrt{\rho_m}}{\sqrt{3}H}\,.
 \ee
 So, the Friedmann equation~(\ref{eq2}) can be recast as
 \be{eq28}
 x^2+y^2-\xi u^2+v^2=1\,.
 \ee
 From the Raychaudhuri equation~(\ref{eq3}) and
 Eqs.~(\ref{eq4}), (\ref{eq20}), (\ref{eq21}), we have
 \be{eq29}
 s\equiv -\frac{\dot{H}}{H^2}=3x^2-\xi su^2+2\sqrt{6}\,\xi xu
 +\frac{3}{2}\gamma v^2\,,
 \ee
 in which $s$ appears in both sides. From Eq.~(\ref{eq29}), it
 is easy to find that
 \be{eq30}
 s=\left(3x^2+2\sqrt{6}\,\xi xu+\frac{3}{2}\gamma v^2\right)
 \left(1+\xi u^2\right)^{-1}.
 \ee
 With the help of Eqs.~(\ref{eq2}), (\ref{eq3}), (\ref{eq20})
 and (\ref{eq21}), the evolution equations~(\ref{eq25}) and
 (\ref{eq26}) can be rewritten as a dynamical system, namely
 \bea
 &&x^\prime=(s-3)x-\sqrt{6}\,\xi u-
 \frac{\kappa V_{,\phi}}{\sqrt{6}H^2}-Q_1\,,\label{eq31}\\
 &&y^\prime=sy+\frac{x}{\sqrt{2}H}
 \frac{V_{,\phi}}{\sqrt{V}}\,,\label{eq32}\\
 &&u^\prime=\sqrt{6}\,x\,,\label{eq33}\\
 &&v^\prime=\left(s-\frac{3}{2}\gamma\right)v
 +Q_2\,,\label{eq34}
 \eea
 where a prime denotes derivative with respect to the so-called
 $e$-folding time $N\equiv\ln a$, and
 \be{eq35}
 Q_1\equiv\frac{\kappa Q}{\sqrt{6}H^2\dot{\phi}}\,,~~~~~~~
 Q_2\equiv\frac{vQ}{2H\rho_m}\,.
 \ee
 On the other hand, the fractional energy densities
 $\Omega_i\equiv (\kappa^2\rho_i)/(3H^2)$ of teleparallel dark
 energy and background matter are given by
 \be{eq36}
 \Omega_\phi=x^2+y^2-\xi u^2\,,~~~~~~~~~
 \Omega_m=v^2\,.
 \ee
 The EoS of teleparallel dark energy reads
 \be{eq37}
 w_\phi\equiv\frac{p_\phi}{\rho_\phi}=\frac{x^2-y^2+\xi u^2-
 \frac{2}{3}\xi s u^2+4\sqrt{\frac{2}{3}}\,\xi xu}{x^2+y^2-
 \xi u^2}\,.
 \ee
 On the other hand, the total EoS is given by
 \be{eq38}
 w_{tot}\equiv\frac{p_{tot}}{\rho_{tot}}=\Omega_\phi w_\phi+
 \Omega_m w_m=x^2-y^2+\xi u^2-\frac{2}{3}\xi s u^2
 +4\sqrt{\frac{2}{3}}\,\xi xu+(\gamma-1)v^2\,.
 \ee
 Eqs.~(\ref{eq31})---(\ref{eq34}) could be an autonomous
 system when the potential $V(\phi)$ and the interaction
 term $Q$ are chosen to be suitable forms. In fact, we will
 consider the models with an exponential and power-law
 potential in the following sections. In each model with
 different potential, we consider four cases with various
 interaction forms between teleparallel dark energy and
 background matter. The first case is the one
 without interaction, i.e., $Q=0$. The other three cases are
 taken as the most familiar interaction terms extensively
 considered in the literature (see
 e.g.~\cite{r26,r27,r28,r29,r30}), namely
 \begin{eqnarray*}
 &{\rm Case~(I)} &Q=0\,,\\
 &{\rm Case~(II)} &Q=\alpha\kappa\rho_m\dot{\phi}\,,\\
 &{\rm Case~(III)} &Q=3\beta H\rho_{tot}\,,\\
 &{\rm Case~(IV)} &Q=3\eta H\rho_m\,,
 \end{eqnarray*}
 where $\alpha$, $\beta$ and $\eta$ are all dimensionless
 constants.\\

%==================== table 1 ====================

 \begin{table}[htbp]
 \begin{center}
 \begin{tabular}{l|c|c}
 \hline\hline & & \\[-4mm]
 ~Label~ & \ Critical Point
 $(\bar{x},\bar{y},\bar{u},\bar{v})$ & \ Existence \\[0.5mm]
 \hline & & \\[-3mm]
 ~E.I.1m~ & ~ 0,\
 $\sqrt{2\left[\xi-\sqrt{\xi\left(
 \xi-\lambda^2\right)}\right]\lambda^{-2}}\,$,
 \ $\left[\xi-\sqrt{\xi\left(
 \xi-\lambda^2\right)}\right]\left(\lambda\xi\right)^{-1}$,
 \ 0 ~ & $\xi\geq\lambda^2$ \\[3mm]
 ~E.I.1p~ & ~ 0,\
 $\sqrt{2\left[\xi+\sqrt{\xi\left(
 \xi-\lambda^2\right)}\right]\lambda^{-2}}\,$,
 \ $\left[\xi+\sqrt{\xi\left(
 \xi-\lambda^2\right)}\right]\left(\lambda\xi\right)^{-1}$,
 \ 0 ~ & ~ $\xi\geq\lambda^2$ or $\xi<0$ \\[3mm]
 ~E.I.2~  & 0,\ 0,\ 0,\ 1 & always \\[1.5mm]
 \hline\hline
 \end{tabular}
 \end{center}
 \caption{\label{tab1} Critical points for the autonomous
 system (\ref{eq40})---(\ref{eq43}) and their corresponding
 existence conditions, for Case~(I) $Q=0$. See text for
 details.}
 \end{table}

%=================================================

\vspace{-6mm}  % used here just for a comfortable typesetting

%============================= section 3 ===================================

\section{Teleparallel dark energy with an
 exponential potential}\label{sec3}

At first, we consider teleparallel dark energy with an
 exponential potential
 \be{eq39}
 V(\phi)=V_0\, e^{-\lambda\kappa\phi}\,,
 \ee
 where $\lambda$ is a dimensionless constant. In this case,
 Eqs.~(\ref{eq31})---(\ref{eq34}) become
 \bea
 &&x^\prime=(s-3)x-\sqrt{6}\,\xi u
 +\sqrt{\frac{3}{2}}\,\lambda y^2-Q_1\,,\label{eq40}\\
 &&y^\prime=sy-\sqrt{\frac{3}{2}}\,\lambda xy\,,\label{eq41}\\
 &&u^\prime=\sqrt{6}\,x\,,\label{eq42}\\
 &&v^\prime=\left(s-\frac{3}{2}\gamma\right)v
 +Q_2\,.\label{eq43}
 \eea
 If $Q$ is given, we can obtain the critical
 points $(\bar{x},\bar{y},\bar{u},\bar{v})$ of
 the above autonomous system by imposing the conditions
 $\bar{x}^\prime=\bar{y}^\prime=\bar{u}^\prime=\bar{v}^\prime=0$.
 Of course, they are subject to the Friedmann
 constraint~(\ref{eq28}), i.e.,
 $\bar{x}^2+\bar{y}^2-\xi\bar{u}^2+\bar{v}^2=1$. On the other
 hand, by definitions in Eq.~(\ref{eq27}), $\bar{x}$,
 $\bar{y}$, $\bar{u}$,  $\bar{v}$ should be real, and
 $\bar{y}\ge 0$, $\bar{v}\ge 0$ are required.

For Case~(I) $Q=0$, the corresponding $Q_1=0$ and $Q_2=0$. In
 this case, there are three critical points, and we present
 these critical points and their corresponding existence
 conditions in Table~\ref{tab1}. From
 Eqs.~(\ref{eq30}), (\ref{eq36}), (\ref{eq37})
 and~(\ref{eq38}), we find that at Points~(E.I.1m) and
 (E.I.1p), $\Omega_\phi=1$, $\Omega_m=0$, $w_\phi=-1$ and
 $w_{tot}=-1$, namely, they are both dark-energy-dominated
 de~Sitter solutions. On the other hand, Point~(E.I.2) is a
 matter-dominated solution. Therefore, there is {\em no}
 scaling solution in Case~(I) $Q=0$. For Case~(II)
 $Q=\alpha\kappa\rho_m \dot{\phi}$,
 the corresponding $Q_1=\sqrt{\frac{3}{2}}\,\alpha v^2$ and
 $Q_2=\sqrt{\frac{3}{2}}\,\alpha xv$. In this case, there are
 four critical points, and we present these critical points
 and their corresponding existence conditions in
 Table~\ref{tab2}. From Eqs.~(\ref{eq30}), (\ref{eq36}),
 (\ref{eq37}) and~(\ref{eq38}), we find that at
 Points~(E.II.2m) and (E.II.2p), $\Omega_\phi=1$,
 $\Omega_m=0$, $w_\phi=-1$ and $w_{tot}=-1$, namely, they are
 both dark-energy-dominated de~Sitter solutions. On the other
 hand, Points~(E.II.1m) and (E.II.1p) are both scaling
 solutions. Thus, they can give the hope to alleviate the
 cosmological coincidence problem. However, their stabilities
 are required in order to be attractors which are necessary to
 this end (see the discussions below). For Case~(III)
 $Q=3\beta H\rho_{tot}$, the corresponding
 $Q_1=\frac{3}{2}\beta x^{-1}$ and
 $Q_2=\frac{3}{2}\beta v^{-1}$. On the other hand, for
 Case~(IV) $Q=3\eta H\rho_m$, the corresponding
 $Q_1=\frac{3}{2}\eta x^{-1}v^2$ and $Q_2=\frac{3}{2}\eta v$.
 Unfortunately, we find that there is {\em no} critical point
 in these two cases.

\vspace{2.5mm}  % used here just for a comfortable typesetting

%==================== table 2 ====================

 \begin{table}[htbp]
 \begin{center}
 \begin{tabular}{l|c|c}
 \hline\hline & & \\[-4mm]
 ~Label~ & \ Critical Point
 $(\bar{x},\bar{y},\bar{u},\bar{v})$ & \ Existence \\[0.5mm]
 \hline & & \\[-3mm]
 ~E.II.1m~ & ~ 0,\ 0,
 \ $\left[-\xi+\sqrt{\xi\left(
 \xi-\alpha^2\right)}\right]\left(\alpha\xi\right)^{-1},$
 \ $\sqrt{2\left[\xi-\sqrt{\xi\left(
 \xi-\alpha^2\right)}\right]\alpha^{-2}}\,$ ~
 & $\xi\geq\alpha^2$ \\[3mm]
 ~E.II.1p~ & ~ 0,\ 0,
 \ $\left[-\xi-\sqrt{\xi\left(
 \xi-\alpha^2\right)}\right]\left(\alpha\xi\right)^{-1},$
 \ $\sqrt{2\left[\xi+\sqrt{\xi\left(
 \xi-\alpha^2\right)}\right]\alpha^{-2}}\,$ ~
 & ~ $\xi\geq\alpha^2$ or $\xi<0$\\[3mm]
 ~E.II.2m~ & ~ 0,\
 $\sqrt{2\left[\xi-\sqrt{\xi\left(
 \xi-\lambda^2\right)}\right]\lambda^{-2}}\,$,
 \ $\left[\xi-\sqrt{\xi\left(
 \xi-\lambda^2\right)}\right]\left(\lambda\xi\right)^{-1}$,
 \ 0 ~ & $\xi\geq\lambda^2$ \\[3mm]
 ~E.II.2p~ & ~ 0,\
 $\sqrt{2\left[\xi+\sqrt{\xi\left(
 \xi-\lambda^2\right)}\right]\lambda^{-2}}\,$,
 \ $\left[\xi+\sqrt{\xi\left(
 \xi-\lambda^2\right)}\right]\left(\lambda\xi\right)^{-1}$,
 \ 0 ~ & ~ $\xi\geq\lambda^2$ or $\xi<0$ \\[3mm]
 \hline\hline
 \end{tabular}
 \end{center}
 \caption{\label{tab2} The same as in Table~\ref{tab1}, except
 for Case~(II) $Q=\alpha\kappa\rho_m \dot{\phi}$.}
 \end{table}

%=================================================

\vspace{1.5mm}  % used here just for a comfortable typesetting

To study the stability of the critical points for
 the autonomous system Eqs.~(\ref{eq40})---(\ref{eq43}), we
 substitute linear perturbations $x\to\bar{x}+\delta x$,
 $y\to\bar{y}+\delta y$, $u\to\bar{u}+\delta u$, and
 $v\to\bar{v}+\delta v$ about the critical point
 $(\bar{x},\bar{y},\bar{u},\bar{v})$ into the autonomous system
 Eqs.~(\ref{eq40})---(\ref{eq43}) and linearize them. Because
 of the Friedmann constraint~(\ref{eq28}), there are only three
 independent evolution equations, namely
 \bea
 &&\delta x^\prime=\left(\bar{s}-3\right)\delta x
 +\bar{x}\delta s-\sqrt{6}\,\xi\delta u+
 \sqrt{6}\,\lambda\bar{y}\delta y-\delta Q_1\,,\label{eq44}\\
 &&\delta y^\prime=\bar{s}\delta y+\bar{y}\delta s
 -\sqrt{\frac{3}{2}}\,\lambda\left(\bar{x}\delta y
 +\bar{y}\delta x\right)\,,\label{eq45}\\
 &&\delta u^\prime=\sqrt{6}\,\delta x\,,\label{eq46}
 \eea
 where
 \bea
 &&\bar{s}=\left[3\bar{x}^2+2\sqrt{6}\,\xi\bar{x}\bar{u}+
 \frac{3}{2}\gamma\left(1-\bar{x}^2-\bar{y}^2+\xi\bar{u}^2
 \right)\right]\left(1+\xi\bar{u}^2\right)^{-1},\label{eq47}\\
 &&\delta s=\left[-2\xi\bar{s}\bar{u}\delta u+6\bar{x}\delta x
 +2\sqrt{6}\,\xi\left(\bar{x}\delta u+\bar{u}\delta x\right)+
 3\gamma\left(\xi\bar{u}\delta u-\bar{x}\delta x
 -\bar{y}\delta y\right)\right]
 \left(1+\xi\bar{u}^2\right)^{-1},\label{eq48}
 \eea
 and $\delta Q_1$ is the linear perturbation coming from $Q_1$.
 The three eigenvalues of the coefficient matrix of
 Eqs.~(\ref{eq44})---(\ref{eq46}) determine the stability of
 the critical point.

For Case~(I) $Q=0$, the corresponding $\delta Q_1=0$. The
 three eigenvalues for Point~(E.I.1m) are the three roots of
 equation (in which $r$ is the unknown quantity)
 \be{eq49}
 \left(r+3\gamma\right)\left\{6\lambda^2 \xi+\left[-\xi
 +\sqrt{\xi\left(\xi-\lambda^2\right)}\right]\times\left[
 6\xi-r\left(3+r\right)\right]\right\}=0\,.
 \ee
 Obviously, the first eigenvalue is $-3\gamma$. The other two
 eigenvalues are complicated and hence we do not give their
 explicit expressions here. We find that if Point~(E.I.1m)
 exists (under condition $\xi\geq\lambda^2$), it is stable.
 Similarly, the three eigenvalues for Point~(E.I.1p) are the
 three roots of equation (in which $r$ is the unknown quantity)
 \be{eq50}
 \left(r+3\gamma\right)\left\{6\lambda^2 \xi-\left[\xi
 +\sqrt{\xi\left(\xi-\lambda^2\right)}\right]\times\left[
 6\xi-r\left(3+r\right)\right]\right\}=0\,.
 \ee
 Again, the first eigenvalue is $-3\gamma$. The other two
 eigenvalues are complicated and hence we do not give their
 explicit expressions here. Unfortunately, we find that
 if Point~(E.I.1p) exists (under condition $\xi\geq\lambda^2$
 or $\xi<0$), it is unstable. Finally, the three eigenvalues
 for Point~(E.I.2) are given by
 \be{eq51}
 \left\{\frac{3\gamma}{2},\ \frac{1}{4}\left[-6+3\gamma-
 \sqrt{9(\gamma-2)^2-96\xi}\right],\ \frac{1}{4}\left[-6+
 3\gamma+\sqrt{9(\gamma-2)^2-96\xi}\right]\right\}.
 \ee
 Because the first eigenvalue $3\gamma/2$ is positive, Point~(E.I.2)
 is unstable.

For Case~(II) $Q=\alpha\kappa\rho_m\dot{\phi}$,
 the corresponding $\delta Q_1=\sqrt{6}\,\alpha
 \left(\xi\bar{u}\delta u-\bar{x}\delta x-
 \bar{y}\delta y\right)$. The three eigenvalues
 for Point~(E.II.1m) are
 \be{eq52}
 \left\{\frac{3\gamma}{2},\ \frac{1}{4}\left(-6+3\gamma-
 \sigma_{-}\right),\ \frac{1}{4}\left(-6+3\gamma+
 \sigma_{-}\right)\right\},
 \ee
 where
 \be{eq53}
 \sigma_{-}\equiv\sqrt{9(\gamma-2)^2-96\sqrt{\xi(\xi-\alpha^2)}}\,.
 \ee
 The three eigenvalues for Point~(E.II.1p) are
 \be{eq54}
 \left\{\frac{3\gamma}{2},\ \frac{1}{4}\left(-6+3\gamma-
 \sigma_{+}\right),\ \frac{1}{4}\left(-6+3\gamma+
 \sigma_{+}\right)\right\},
 \ee
 where
 \be{eq55}
 \sigma_{+}\equiv\sqrt{9(\gamma-2)^2+96\sqrt{\xi(\xi-\alpha^2)}}\,.
 \ee
 Since their first eigenvalue $3\gamma/2$ is positive,
 Points~(E.II.1m) and (E.II.1p) are both unstable. Therefore,
 although they are scaling solutions, however, they are not
 attractors and hence cannot alleviate the cosmological coincidence
 problem. On the other hand, the three eigenvalues for
 Point~(E.II.2m) are the three roots of Eq.~(\ref{eq49}). So,
 if Point~(E.II.2m) exists (under condition $\xi\geq\lambda^2$), it
 is stable. Similarly, the three eigenvalues for Point~(E.II.2p) are
 the three roots of Eq.~(\ref{eq50}). So, if
 Point~(E.II.2p) exists (under condition $\xi\geq\lambda^2$
 or $\xi<0$), it is unstable.

Since in both Case~(III) $Q=3\beta H\rho_{tot}$ and Case~(IV)
 $Q=3\eta H\rho_m$ there is no critical point, we need not
 perform the stability analysis for them.

So, for teleparallel dark energy with an exponential potential,
 in Case~(I) $Q=0$ there is only one attractor~(E.I.1m) which
 is a dark-energy-dominated de~Sitter solution, and in
 Case~(II) $Q=\alpha\kappa\rho_m\dot{\phi}$ there is only one
 attractor~(E.II.2m) which is also a dark-energy-dominated de~Sitter
 solution. No scaling attractor has been found unfortunately.

%============================= section 4 ===================================

\section{Teleparallel dark energy with a power-law
 potential}\label{sec4}

Due to the failure in the models with an exponential potential,
 we turn to teleparallel dark energy with a power-law potential
 \be{eq56}
 V(\phi)=V_0\left(\kappa\phi\right)^n\,,
 \ee
 where $n$ is a dimensionless constant. In this case,
 Eqs.~(\ref{eq31})---(\ref{eq34}) become
 \bea
 &&x^\prime=(s-3)x-\sqrt{6}\,\xi u
 -\sqrt{\frac{3}{2}}\,n y^2 u^{-1}-Q_1\,,\label{eq57}\\
 &&y^\prime=sy+\sqrt{\frac{3}{2}}\,n xyu^{-1}\,,\label{eq58}\\
 &&u^\prime=\sqrt{6}\,x\,,\label{eq59}\\
 &&v^\prime=\left(s-\frac{3}{2}\gamma\right)v
 +Q_2\,.\label{eq60}
 \eea
 If $Q$ is given, we can obtain the critical
 points $(\bar{x},\bar{y},\bar{u},\bar{v})$ of
 the above autonomous system by imposing the conditions
 $\bar{x}^\prime=\bar{y}^\prime=\bar{u}^\prime=\bar{v}^\prime=0$.
 Of course, they are subject to the Friedmann
 constraint~(\ref{eq28}), i.e.,
 $\bar{x}^2+\bar{y}^2-\xi\bar{u}^2+\bar{v}^2=1$. On the other
 hand, by definitions in Eq.~(\ref{eq27}), $\bar{x}$,
 $\bar{y}$, $\bar{u}$,  $\bar{v}$ should be real, and
 $\bar{y}\ge 0$, $\bar{v}\ge 0$ are required.

%==================== table 3 ====================

 \begin{table}[tbp]
 \begin{center}
 \begin{tabular}{l|c|c}
 \hline\hline & & \\[-4mm]
 ~Label~ & \ Critical Point
 $(\bar{x},\bar{y},\bar{u},\bar{v})$ & \ Existence \\[0.5mm]
 \hline & & \\[-3mm]
 ~P.I.1m~ & ~ 0,
 \ $\sqrt{\frac{2}{n+2}}\,$,
 \ $-\sqrt{\frac{n}{-\xi (n+2)}}\,$, \ 0 ~
 & ~ (a) $\xi<0$ and $n\geq 0$
 ~or~ (b) $\xi>0$ and $-2<n\leq 0\,$\\[3mm]
 ~P.I.1p~ & ~ 0,
 \ $\sqrt{\frac{2}{n+2}}\,$,
 \ $\sqrt{\frac{n}{-\xi (n+2)}}\,$, \ 0 ~
 & ~ (a) $\xi<0$ and $n\geq 0$
 ~or~ (b) $\xi>0$ and $-2<n\leq 0\,$\\[3mm]
 \hline\hline
 \end{tabular}
 \end{center}
 \caption{\label{tab3} Critical points for the autonomous
 system (\ref{eq57})---(\ref{eq60}) and their corresponding
 existence conditions, for Case~(I) $Q=0$. See text for
 details.}
 \end{table}

%=================================================

\vspace{4mm}  % used here just for a comfortable typesetting

%==================== table 4 ====================

 \begin{table}[htbp]
 \begin{center}
 \begin{tabular}{l|c|c}
 \hline\hline & & \\[-4mm]
 ~Label~ & \ Critical Point
 $(\bar{x},\bar{y},\bar{u},\bar{v})$ & \ Existence \\[0.5mm]
 \hline & & \\[-3mm]
 ~P.II.1m~ & ~ 0,\ 0,
 \ $\left[-\xi+\sqrt{\xi\left(
 \xi-\alpha^2\right)}\right]\left(\alpha\xi\right)^{-1},$
 \ $\sqrt{2\left[\xi-\sqrt{\xi\left(
 \xi-\alpha^2\right)}\right]\alpha^{-2}}\,$ ~
 & $\xi\geq\alpha^2$ \\[3mm]
 ~P.II.1p~ & ~ 0,\ 0,
 \ $\left[-\xi-\sqrt{\xi\left(
 \xi-\alpha^2\right)}\right]\left(\alpha\xi\right)^{-1},$
 \ $\sqrt{2\left[\xi+\sqrt{\xi\left(
 \xi-\alpha^2\right)}\right]\alpha^{-2}}\,$ ~
 & ~ $\xi\geq\alpha^2$ or $\xi<0$\\[3mm]
 ~P.II.2m~ & ~ 0,
 \ $\sqrt{\frac{2}{n+2}}\,$,
 \ $-\sqrt{\frac{n}{-\xi (n+2)}}\,$, \ 0 ~
 & ~ the same as for Point (P.I.1m)\\[3mm]
 ~P.II.2p~ & ~ 0,
 \ $\sqrt{\frac{2}{n+2}}\,$,
 \ $\sqrt{\frac{n}{-\xi (n+2)}}\,$, \ 0 ~
 & ~ the same as for Point (P.I.1p)\\[3mm]
 \hline\hline
 \end{tabular}
 \end{center}
 \caption{\label{tab4} The same as in Table~\ref{tab3}, except
 for Case~(II) $Q=\alpha\kappa\rho_m \dot{\phi}$.}
 \end{table}

%=================================================

\vspace{2mm}  % used here just for a comfortable typesetting

For Case~(I) $Q=0$, the corresponding $Q_1=0$ and $Q_2=0$. In
 this case, there are two critical points, and we present
 these critical points and their corresponding existence
 conditions in Table~\ref{tab3}. From
 Eqs.~(\ref{eq30}), (\ref{eq36}), (\ref{eq37})
 and~(\ref{eq38}), we find that at Points~(P.I.1m) and
 (P.I.1p), $\Omega_\phi=1$, $\Omega_m=0$, $w_\phi=-1$ and
 $w_{tot}=-1$, namely, they are both dark-energy-dominated
 de~Sitter solutions. Therefore, there is {\em no}
 scaling solution in Case~(I) $Q=0$. For Case~(II)
 $Q=\alpha\kappa\rho_m \dot{\phi}$,
 the corresponding $Q_1=\sqrt{\frac{3}{2}}\,\alpha v^2$ and
 $Q_2=\sqrt{\frac{3}{2}}\,\alpha xv$. In this case, there are
 four critical points, and we present these critical points
 and their corresponding existence conditions in
 Table~\ref{tab4}. From
 Eqs.~(\ref{eq30}), (\ref{eq36}), (\ref{eq37})
 and~(\ref{eq38}), we find that at Points~(P.II.2m) and
 (P.II.2p), $\Omega_\phi=1$, $\Omega_m=0$, $w_\phi=-1$ and
 $w_{tot}=-1$, namely, they are both dark-energy-dominated
 de~Sitter solutions. On the other
 hand, Points~(P.II.1m) and (P.II.1p) are both scaling
 solutions. So, they can give the hope to alleviate the
 cosmological coincidence problem. However, their stabilities
 are required in order to be attractors which are necessary to
 this end (see the discussions below). For Case~(III)
 $Q=3\beta H\rho_{tot}$, the corresponding
 $Q_1=\frac{3}{2}\beta x^{-1}$ and
 $Q_2=\frac{3}{2}\beta v^{-1}$. On the other hand, for
 Case~(IV) $Q=3\eta H\rho_m$, the corresponding
 $Q_1=\frac{3}{2}\eta x^{-1}v^2$ and $Q_2=\frac{3}{2}\eta v$.
 Unfortunately, we find that there is {\em no} critical point
 in these two cases.

To study the stability of the critical points for
 the autonomous system Eqs.~(\ref{eq57})---(\ref{eq60}), we
 substitute linear perturbations $x\to\bar{x}+\delta x$,
 $y\to\bar{y}+\delta y$, $u\to\bar{u}+\delta u$, and
 $v\to\bar{v}+\delta v$ about the critical point
 $(\bar{x},\bar{y},\bar{u},\bar{v})$ into the autonomous system
 Eqs.~(\ref{eq57})---(\ref{eq60}) and linearize them. Because
 of the Friedmann constraint~(\ref{eq28}), there are only three
 independent evolution equations, namely
 \bea
 &&\delta x^\prime=\left(\bar{s}-3\right)\delta x
 +\bar{x}\delta s-\sqrt{6}\,\xi\delta u-
 \sqrt{\frac{3}{2}}\,n\bar{y}\bar{u}^{-1}\left(2\delta y-
 \bar{y}\bar{u}^{-1}\delta u\right)-\delta Q_1\,,\label{eq61}\\
 &&\delta y^\prime=\bar{s}\delta y+\bar{y}\delta s
 +\sqrt{\frac{3}{2}}\,n\bar{u}^{-1}\left(\bar{x}\delta y
 +\bar{y}\delta x-\bar{x}\bar{y}\bar{u}^{-1}\delta u\right)\,,
 \label{eq62}\\
 &&\delta u^\prime=\sqrt{6}\,\delta x\,,\label{eq63}
 \eea
 where $\delta Q_1$ is the linear perturbation coming from
 $Q_1$, and $\bar{s}$, $\delta s$ are given in
 Eqs.~(\ref{eq47}) and~(\ref{eq48}). The three eigenvalues of
 the coefficient matrix of Eqs.~(\ref{eq61})---(\ref{eq63})
 determine the stability of the critical point.

For Case~(I) $Q=0$, the corresponding $\delta Q_1=0$. The
 three eigenvalues for both Points~(P.I.1m) and~(P.I.1p) are
 given by
 \be{eq64}
 \left\{-3\gamma,\ \frac{1}{2}\left[-3-\sqrt{9-24\xi (n+2)}\right],
 \ \frac{1}{2}\left[-3+\sqrt{9-24\xi (n+2)}\right]\right\}\,.
 \ee
 So, both Points~(P.I.1m) and~(P.I.1p) exist and are stable
 under condition $\xi>0$ and $-2<n\leq 0$.

For Case~(II) $Q=\alpha\kappa\rho_m\dot{\phi}$,
 the corresponding $\delta Q_1=\sqrt{6}\,\alpha
 \left(\xi\bar{u}\delta u-\bar{x}\delta x-
 \bar{y}\delta y\right)$. The three eigenvalues
 for Points~(P.II.1m) and~(P.II.1p) are given by
 Eq.~(\ref{eq52}) and~(\ref{eq54}), respectively. Since their
 first eigenvalue $3\gamma/2$ is positive, Points~(P.II.1m)
 and (P.II.1p) are both unstable. So, although they are scaling
 solutions, however, they are not attractors and hence cannot
 alleviate the cosmological coincidence problem. On the other
 hand, the three eigenvalues for both Points~(P.II.2m)
 and~(P.II.2p) are given by Eq.~(\ref{eq64}). So, they exist
 and are stable under condition $\xi>0$ and $-2<n\leq 0$.

Since in both Case~(III) $Q=3\beta H\rho_{tot}$ and Case~(IV)
 $Q=3\eta H\rho_m$ there is no critical point, we need not
 perform the stability analysis for them.

So, for teleparallel dark energy with a power-law potential,
 in Case~(I) $Q=0$ there are two attractors~(P.I.1m) and
 (P.I.1p) which are both dark-energy-dominated de~Sitter
 solutions, and in Case~(II) $Q=\alpha\kappa\rho_m\dot{\phi}$
 there are two attractors~(P.II.2m) and (P.II.2p) which are
 also dark-energy-dominated de~Sitter solutions. Again, no
 scaling attractor has been found unfortunately.

%============================= section 5 ===================================

\section{Concluding remarks}\label{sec5}

Recently, Geng~{\it et~al.}~\cite{r15} proposed to allow a
 non-minimal coupling between quintessence and gravity in the
 framework of teleparallel gravity, motivated by the similar
 one in the framework of GR. They found that this non-minimally
 coupled quintessence in the framework of teleparallel gravity
 has a richer structure, and named it ``teleparallel dark
 energy''~\cite{r15}. In the present work, we note that there
 might be a deep and unknown connection between teleparallel
 dark energy and Elko spinor dark energy. Motivated by this
 observation and the previous results of Elko spinor dark
 energy~\cite{r18,r22}, we try to study the dynamics of
 teleparallel dark energy. We find that there exist only some
 dark-energy-dominated de~Sitter attractors. Unfortunately,
 no scaling attractor has been found, even when we allow the
 possible interaction between teleparallel dark energy and
 matter. However, we note that $w$ at the critical points is
 in agreement with observations (in particular, the fact
 that $w=-1$ independently of $\xi$ is a great advantage).

Some remarks are in order. Firstly, in the present work we have
 chosen some particular potentials $V(\phi)$ and interaction
 forms $Q$. So, it is still possible to find some suitable and
 delicate potentials $V(\phi)$ and interaction forms $Q$ to
 obtain the scaling attractors of the most general dynamical
 system~(\ref{eq31})---(\ref{eq34}), and hence the hope to
 alleviate the cosmological coincidence problem still exists,
 although this is a fairly hard task (see e.g.~\cite{r22}). Of
 course, there also might be other smart methods, different
 from the usual method used in most of dark energy models,
 to alleviate the cosmological coincidence problem (see
 e.g.~\cite{r31}). Secondly, we would like to briefly discuss
 the most general case. From the most general Eq.~(\ref{eq33}),
 we have $\bar{x}=0$ at the critical points. From
 Eq.~(\ref{eq28}), it is easy to see that
 $1+\xi\bar{u}^2=\bar{y}^2+\bar{v}^2\ge 0$. Therefore, from
 Eq.~(\ref{eq30}), we find that $\bar{s}>0$ when
 $\bar{v}\not=0$ (i.e., $\Omega_m\not=0$, scaling solution),
 whereas $\bar{s}=0$ when $\bar{v}=0$ (i.e.,~$\Omega_m=0$,
 dark-energy-dominated solution). From Eq.~(\ref{eq37}), we have
 $\bar{w}_\phi=-1-(2/3)\xi\bar{s}\bar{u}^2/(\bar{y}^2-\xi\bar{u}^2)$.
 From Eq.~(\ref{eq36}), $\bar{y}^2-\xi\bar{u}^2=\Omega_\phi\geq 0$.
 Therefor, in the case of $\bar{s}=0$ when $\bar{v}=0$ (i.e.,
 $\Omega_m=0$, dark-energy-dominated solution), we always have
 $\bar{w}_\phi=-1$ and then $\bar{w}_{tot}=-1$, regardless of $\xi$.
 In the case of $\bar{s}>0$ when $\bar{v}\not=0$ (i.e.,
 $\Omega_m\not=0$, scaling solution), if $\xi\leq 0$ we have
 $\bar{w}_\phi\geq -1$, and if $\xi>0$ we obtain $\bar{w}_\phi<-1$.
 Therefore, teleparallel dark energy being quintessence-like
 ($w_\phi>-1$) or phantom-like ($w_\phi<-1$) at the scaling
 attractors (if any) heavily depends on the sign of $\xi$.
 If $\xi\leq 0$, our universe can avoid the fate of big rip.
 Thirdly, strictly speaking, for $\xi>0$ the phase space is not
 compact, namely one could also investigate possible critical
 points at infinity (i.e., when $\phi$ and $\dot{\phi}$
 diverge)~\cite{r35}. Since this issue has been discussed in detail
 by Xu~{\it et al.}~\cite{r36} (which appeared in arXiv after our
 submission), we do not consider it any more. Finally, as mentioned
 by Geng~{\it et al.}~\cite{r15}, teleparallel dark energy has some
 interesting features, and we consider that this model still
 deserves further investigations.

%============================= acknowledgments ===================================

\section*{ACKNOWLEDGMENTS}

We thank the anonymous referee for quite useful comments and
 suggestions, which help us to improve this work. We are
 grateful to Professors Rong-Gen~Cai and Shuang~Nan~Zhang
 for helpful discussions. We thank Minzi~Feng, as well as
 Hao-Yu~Qi, Xiao-Peng~Ma, Xiao-Jiao~Guo, Long-Fei~Wang, for
 kind help and discussions. This work was supported in part by
 NSFC under Grants No.~11175016 and No.~10905005, as well as
 NCET under Grant No.~NCET-11-0790, and the Fundamental
 Research Fund of Beijing Institute of Technology. It was
 completed during the period of KITPC Program ``Dark Matter
 and New Physics'', which was supported in part by the Project
 of Knowledge Innovation Program (PKIP) of Chinese Academy of
 Sciences, Grant No.~KJCX2.YW.W10.

\renewcommand{\baselinestretch}{1.28}

%============================= references ==================================

\end{document}